\documentclass[english,12pt]{article}
          \usepackage{babel}
          \usepackage[T1]{fontenc}
          \usepackage[latin2]{inputenc}
          \usepackage{pslatex}
\begin{document}

\title{\bf The 2003 and 2005 superhumps in V1113 Cygni.}
\author{K. ~B~\k{a}~k~o~w~s~k~a$^1$, ~A. ~O~l~e~c~h$^2$, ~K. ~Z~{\l}~o~c~z~e~w~s~k~i$^2$ \\
~and~ ~M. ~W~i~\'s~n~i~e~w~s~k~i$^2$}
\date{$^1$ Astronomical Observatory, Adam Mickiewicz University, Ul. S{\l}oneczna 36, 60-286 Pozna\'{n},
Poland\\ {\tt e-mail: bakowska@lab.astro.amu.edu.pl}\\
$^2$ Nicolaus Copernicus Astronomical Center,
Polish Academy of Sciences,
Ul.~Bartycka~18, 00-716~Warszawa, Poland\\
{\tt e-mail: (olech,kzlocz,mwisniew)@camk.edu.pl}\\
~\\
}
\maketitle

\begin{abstract}  

We report CCD photometry of the cataclysmic variable V1113 Cygni. During
two campaigns, lasting from May to August 2003 and from March to June
2005, we recorded two superoutburst. In the obtained light curves we
detected clear superhumps with a mean period  $P_{\rm sh}=0.07891(3)$
days ($113.63\pm0.04$ min). That fact confirms that the star is a member
of SU UMa class of dwarf novae. During the first observed superoutburst
the superhump period was decreasing with an enormous rate of $\dot P =
-4.5(8)\times 10^{-4}$ which is one of the highest values ever observed
in SU UMa systems.


\end{abstract}

\section{Introduction}

Cataclysmic variable stars are close binary systems containing white
dwarf (the primary) and red dwarf (the secondary). In those systems the
secondary loses its mass through the inner Lagrangian point in the
Roche-lobe and the primary star accretes it (Warner 1995).

SU UMa class of dwarf novae is one of the subclasses of cataclysmic
variable stars. SU UMa-type stars show two types of outbursts: short
(outbursts) and long (superoutbursts). Superoutbursts are about one
magnitude brighter than ordinary outbursts and have a duration a few
days longer than the outbursts. In the light curves of superoutbursts
one can see periodic light oscillations (superhumps) with a period a
few percent longer than the orbital period of the binary system. The
thermal-tidal instability model proposed by Osaki (1996) is now known
as the best explanation of the mechanism of the superoutbursts and
superhumps (for a review see Warner 1995, Osaki 1996, Hellier
2001). However, in recent years some serious criticism concerning both
the nature of superoutburst and nature of superhumps has appeared
(Schreiber and Lasota 2007, Smak 2004, 2008, 2009).
 
V1113 Cyg was discovered as a dwarf nova by Hoffmeister (1966), but
received little attention for almost 30 years. Results of Kato et al.
(1995) have shown that the star is SU UMa-type with superhumps'
period of $0.0792(\pm0.0001)$ d. They also reported that V1113 Cyg has a
large ($\sim$ 6 mag) superoutburst amplitude.

In the second paper by Kato et al. (2001) the detection of 30 outburst
from July 1994 to May 2001 is presented. They noticed low number of
outbursts (only about two normal outbursts per each supercycle) and high
number of superoutbursts (12 from 30 detected). They suspected variation
of the mass transfer rate. Because results are inconclusive further
observations were needed.

Kato et al. (2009) reanalyzed the observations of Kato et al. (1995)
and obtained new observations during the 2008 superoutburst. They
reported decreasing superhump periods in both superoutburts with the
rates of $-19.2(6.8)\times 10^{-5}$ and $-5.2(4.7)\times 10^{-5}$,
respectively.

\section{Observations and data reduction} 

Observations of the V1113 Cyg were made during 34 nights (11 nights
between May 5 and August 28 in 2003 and 23 nights between January 17 and
June 30 in 2005). All data were collected at the Ostrowik station of the
Warsaw University Observatory.  V1113 Cyg was observed by 60-cm
Cassegrain telescope equipped with a Tektronics TK512CB back-illuminated
CCD camera. The scale of the camera was 0.76"/pixel providing a
6.5'$\times$6.5' field of view. The full description of the telescope
and camera was given by Udalski and Pych (1992).

We monitored the star in "white light" to have possibility of making
reliable photometry in quiescence and to minimize guiding errors.
The exposure times varied from 19 to 400 seconds depending on the sky
transparency, seeing and brightness of the object.

In Table 1 we present a full journal of our CCD observations of V1113
Cyg. The star was monitored during 48.71 hours. We obtained 1436 useful
exposures. 

\begin{table}[!ht]
{\tiny
\caption{Journal of the CCD observations of V1113 Cygni.}
\begin{center}
\begin{tabular}{|l|c|c|r|r|}
\hline
\hline
Date   & No. of & HJD Start & HJD End & Length \\
       & frames & 2450000. + & 2450000. + & [hr]~ \\
\hline
2003 May 05/06 & 130 & 2765.43110 & 2765.56406 & 3.19\\
2003 May 06/07 & 158 & 2766.44305 & 2766.57682 & 3.21\\
2003 May 07/08 & 172 & 2767.41675 & 2767.56763 & 3.62\\
2003 May 08/09 & 387 & 2768.32876 & 2768.56835 & 5.75\\
2003 May 10/11 & 110 & 2770.46436 & 2770.56406 & 2.40\\
2003 May 12/13 &  79 & 2772.41856 & 2772.56604 & 3.54\\
2003 May 16/17 &  12 & 2776.33366 & 2776.40147 & 1.63\\
2003 May 17/18 &  34 & 2777.37087 & 2777.54984 & 4.30\\
2003 Aug 24/25 &  40 & 2876.49714 & 2876.54793 & 1.22\\
2003 Aug 25/26 &   3 & 2877.39452 & 2877.39761 & 0.07\\
2003 Aug 27/28 &   3 & 2879.45057 & 2879.45727 & 0.16\\
2005 Jan 17/18 &   9 & 3388.16347 & 3388.18044 & 0.41\\
2005 Mar 31/01 &   7 & 3460.61779 & 3469.63519 & 0.42\\
2005 Apr 03/04 &   5 & 3463.59457 & 3463.60723 & 0.30\\
2005 Apr 05/06 &   8 & 3465.56569 & 3465.59401 & 0.68\\
2005 Apr 29/30 &   9 & 3489.55904 & 3489.58260 & 0.57\\
2005 May 07/08 &   6 & 3498.46126 & 3498.47557 & 0.34\\
2005 May 10/11 &   5 & 3501.44756 & 3501.46025 & 0.30\\
2005 May 20/21 &   6 & 3511.45248 & 3511.46966 & 0.41\\
2005 May 21/22 &  10 & 3512.47513 & 3512.49981 & 0.59\\
2005 May 27/28 &   5 & 3518.44534 & 3518.46054 & 0.36\\
2005 May 28/29 &   5 & 3519.44329 & 3519.46939 & 0.41\\
2005 May 29/30 &   6 & 3520.44504 & 3520.46031 & 0.37\\
2005 May 31/01 &  10 & 3522.46051 & 3522.47862 & 0.43\\
2005 Jun 05/06 &   3 & 3527.42791 & 3527.43599 & 0.19\\
2005 Jun 08/09 &   3 & 3530.41011 & 3530.42070 & 0.25\\
2005 Jun 09/10 &   3 & 3531.49101 & 3531.49923 & 0.20\\
2005 Jun 13/14 &  47 & 3535.38920 & 3535.54075 & 3.64\\
2005 Jun 23/24 &  59 & 3545.38505 & 3545.52139 & 3.72\\
2005 Jun 24/25 &  68 & 3546.36543 & 3546.53387 & 4.04\\
2005 Jun 25/26 &  23 & 3547.40972 & 3547.46837 & 1.41\\
2005 Jun 27/28 &   4 & 3549.39505 & 3549.40398 & 0.21\\
2005 Jun 29/30 &   4 & 3551.39310 & 3551.40208 & 0.22\\
2005 Jun 30/01 &   3 & 3552.40636 & 3552.41280 & 0.15\\

\hline
TOTAL & 1436 & & & 48.71 \\
\hline
\hline
\end{tabular}
\end{center}}
\end{table}

All data were de-biased, dark current subtracted and flat-fielded using
a standard procedure based on the IRAF\footnote{IRAF is distributed by
the National Optical Astronomy Observatory, which is operated by the
Association of Universities for Research in Astronomy, Inc., under a
cooperative agreement with the National Science Foundation.} package.
Profile photometry was derived using DAOPHOTII package (Stetson 1987). 

We obtained the relative unfiltered magnitudes of V1113 Cyg by taking
the difference between the magnitude of the variable star and the intensity averaged magnitude of two comparison stars. In Fig. 1 V1113 Cyg is marked as V1 and two
comparison stars are marked as C1 and C2, respectively. The typical accuracy
of our measurments varied between 0.014 and 0.13 mag depending on the
weather conditions and brightness of the object. The median value of the
photometric errors was 0.027 mag.

\vspace*{10.8cm}
\includegraphics{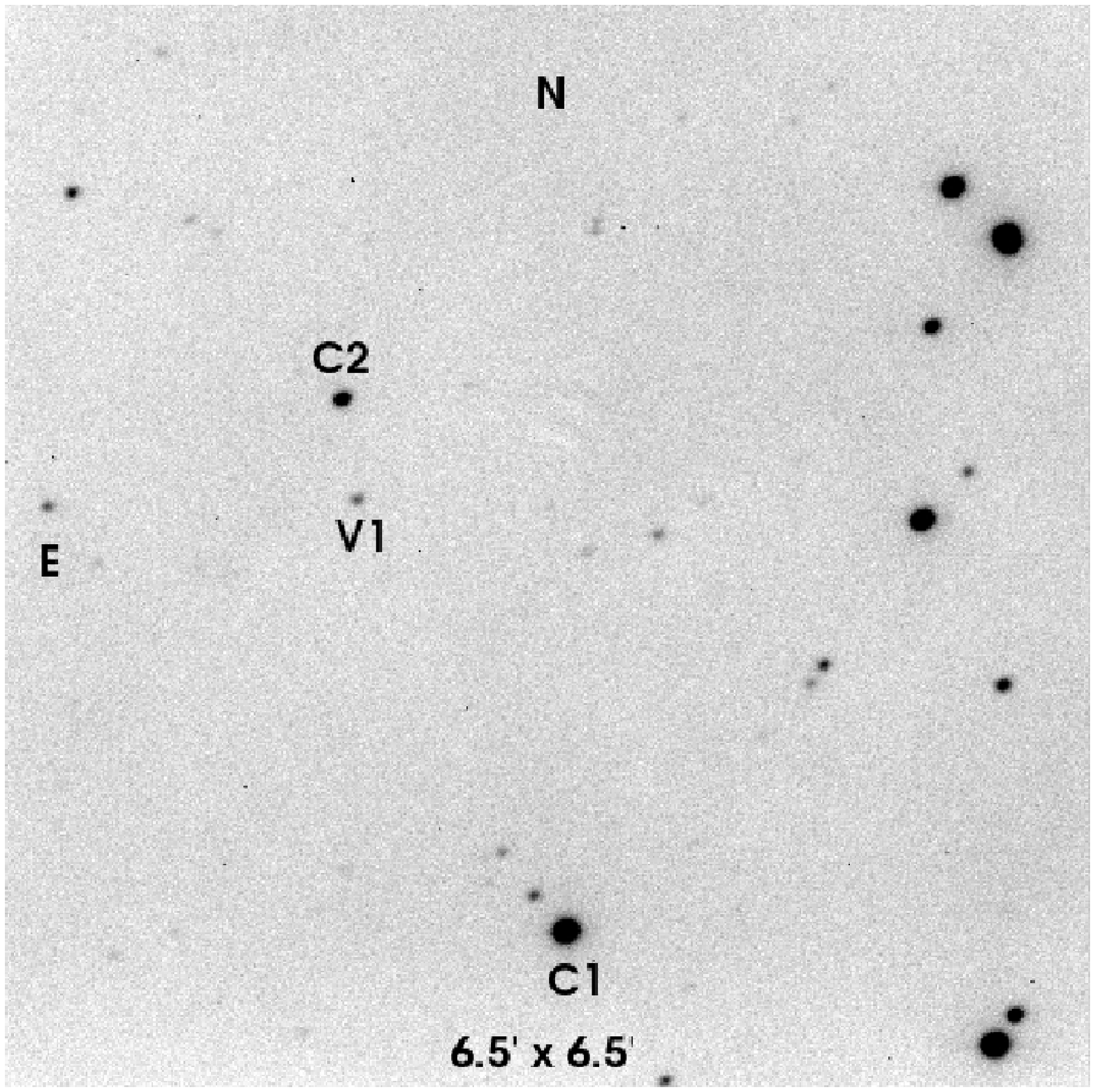}

   \begin{figure}[h]
      \caption {Finding chart of V1113 Cyg. The field of view is about 6.5'$\times$6.5'. 
The variable is marked as V1, comparison stars as C1 and C2, respectively. 
North is up, East is left.}
   \end{figure}  
\section{Light curves}

Figure 2 presents the photometric behavior of V1113 Cyg observed from
May to August in 2003 (upper frame) and from January to June in 2005
(lower frame). Relative magnitudes of the variable were transformed to
the visual scale using the magnitudes of V1113 Cyg estimated by amateur
astronomers and published in AAVSO. Due to this fact it is a rough
estimate, which is used only to show the general behavior of the star.
The true $V$-band brightness of the V1113 Cyg may differ even by about
$\sim$0.3 mag from that shown in Fig. 2. 

\vspace*{16.8cm}

\includegraphics{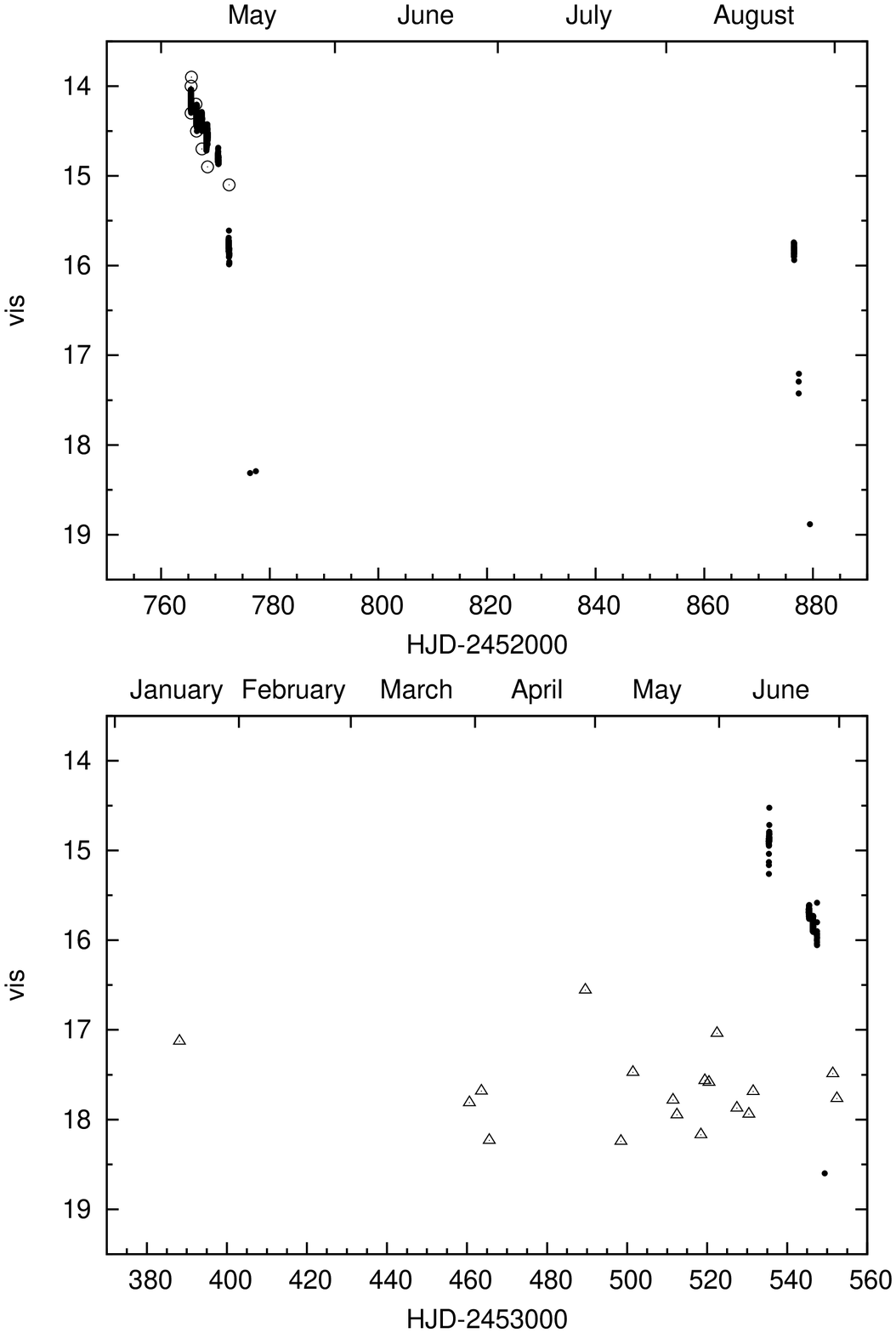}

   \begin{figure}[h]
      \caption {The general photometric behavior of V1113 Cyg during the 2003 
(upper frame) and the 2005 (lower frame) campaigns. On the charts our CCD 
observations are described as black dots. Visual estimates collected by 
AAVSO observers are shown as open circles. The open triangles correspond 
to the "fainter than" the faintest star detected on our CCD frames, which we 
were able to estimate during profile photometry process.}
   \end{figure}

In May 2003 we detected V1113 Cyg during superoutburst. On the first
day of observation - May 5/6, 2003 the star reached a mean magnitude
of 14.2 mag. Systematic decrease of the brightness with a rate of 0.15
mag per day suggest that the star was not caught at the beginning of
its superoutburst, which exact time is difficult to estimate due to
lack of observations. In August 2003 we detected only three days of
increased brightness of the V1113 Cyg. It could be the second
superoutburst, but the normal outburst is also possible. In June 2005
we observed the subsequent superoutburst of the variable. Our
observation from June 9/10 caught the star around minimum light, the
variable was fainter than 17.5 mag. In June 13/14 V1113 Cyg reached a
mean magnitude of 14.89 mag. The brightness of the star was declining
with a rate of 0.13 mag/day during three succesive nights of
superoutburst (June 23/24, 24/25, 25/26). \\

\section{Superhumps}

\subsection{{\sc anova} analysis}

Figure 3 shows the light curves of V1113 Cyg during six nights of the
May 2003 superoutburst. The superhumps are clearly visible in each run
in four consecutive nights. Amplitudes are around 0.3-0.2 mag from the
first to four night, respectively. In May 10/11 the superhump is poorly
visible and due to this fact no amplitude was estimated.

\vspace*{15.0cm}

\includegraphics{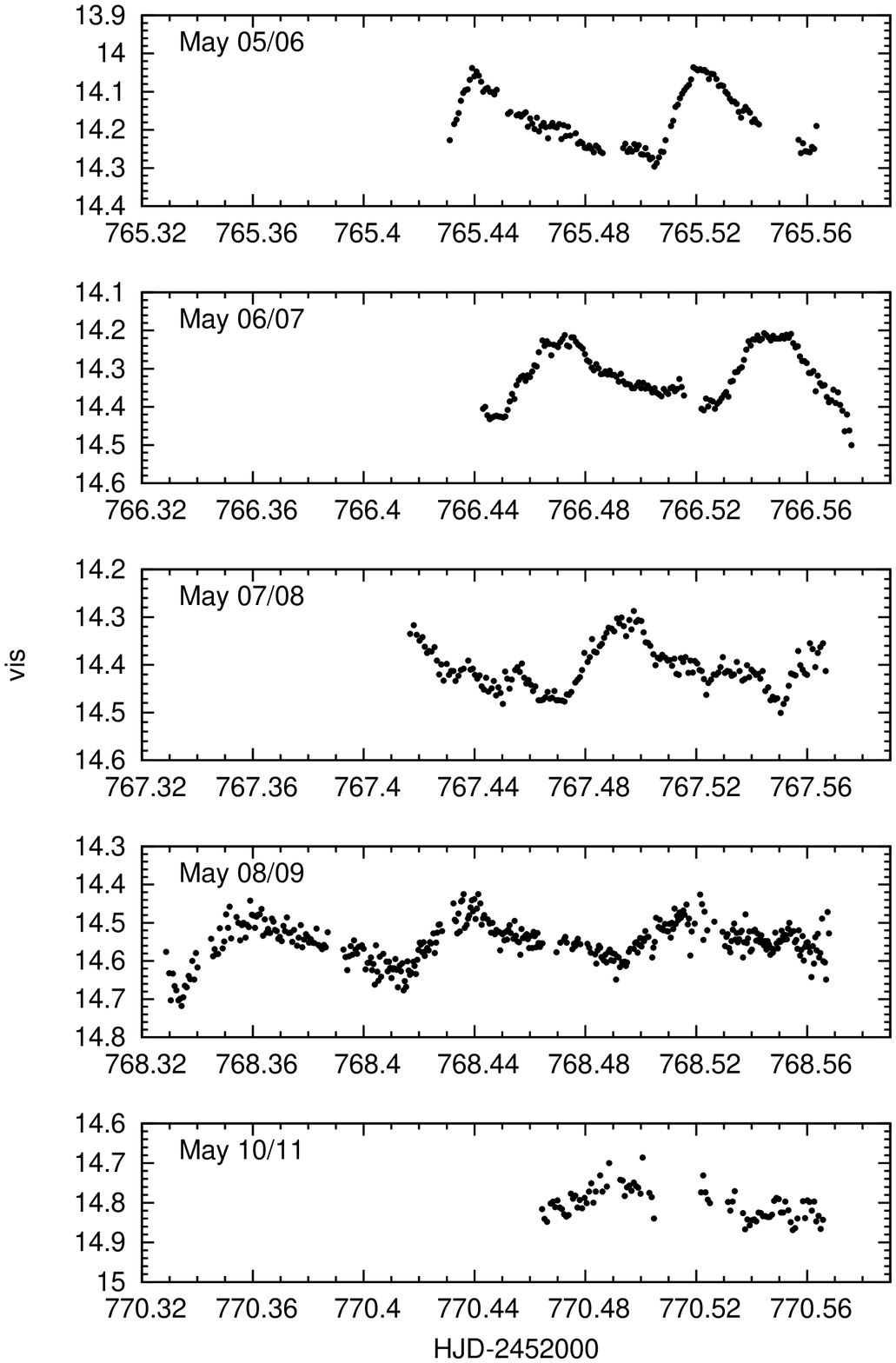}

   \begin{figure}[h]
      \caption {The light curves of V1113 Cyg during its 2003 May superoutburst.}
   \end{figure}

From each light curve of V1113 Cyg in May 2003 and in June 2005
superoutbursts we removed the longer declining trend using first or
second order polynomial and analyzed them using ANOVA statistics with
one or two harmonic Fourier series (Schwarzenberg-Czerny 1996). The
resulting periodograms are shown in Fig. 4. The most prominent peaks are
found at a frequencies of $f_1=12.673(20)$ and $f_2=12.731(30)$ c/d for the
data from 2003 and 2005, respectively. Those frequencies correspond to a
period $P_{\rm 1}=0.07891(12)$ days ($113.63\pm0.17$ min)  and $P_{\rm
2}=0.07855(19)$ days ($113.11\pm0.27$ min) for superhumps in 2003 and 2005,
respectively. \\

\vspace*{17.8cm}

\includegraphics{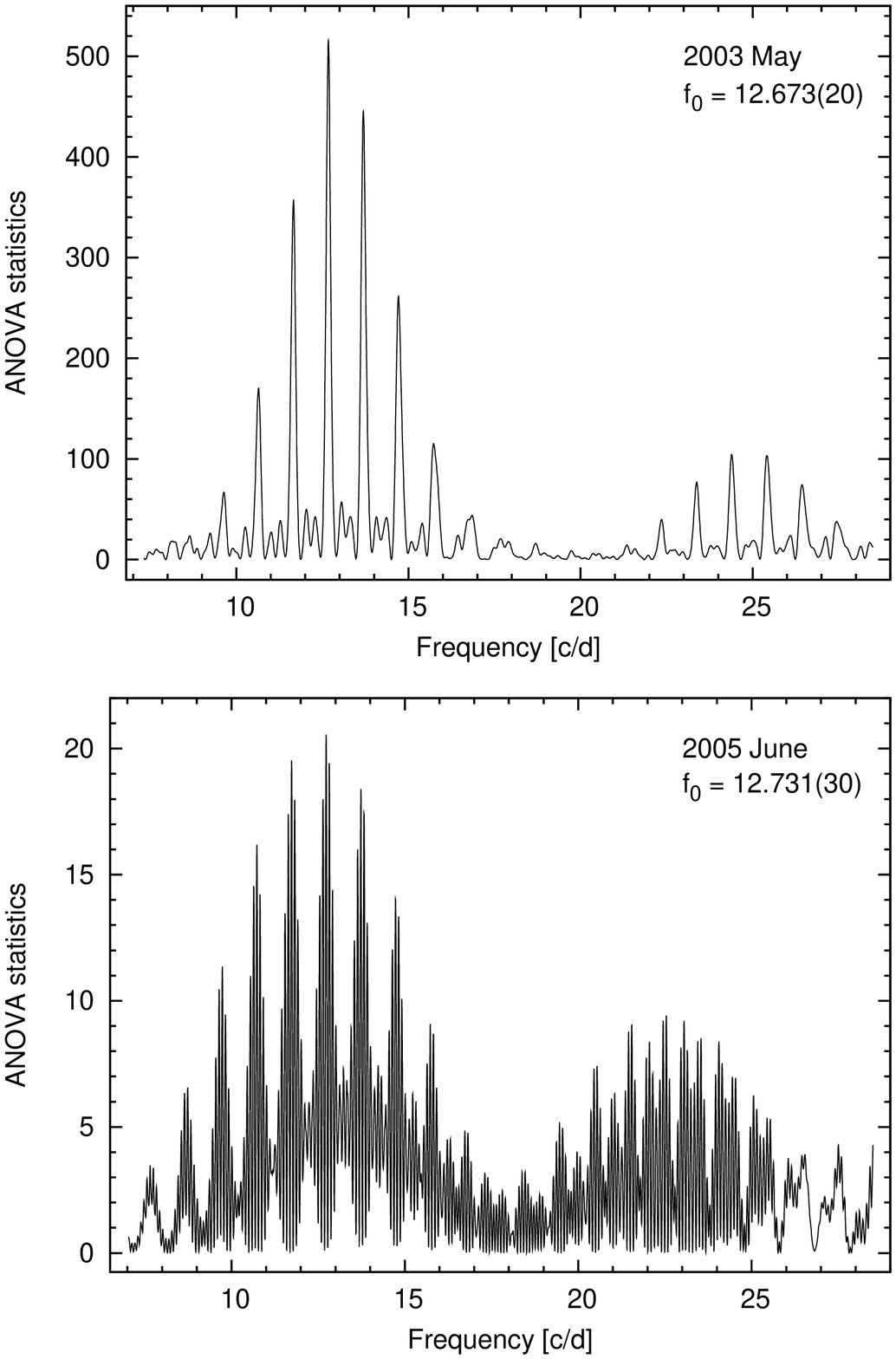}

   \begin{figure}[h]
      \caption {The ANOVA power spectrum of the light curves of V1113 Cyg 
from its 2003 May (upper frame) and 2005 June (lower frame) superoutbursts.}
   \end{figure}  

\subsection{The $O - C$ analysis}

One of the most popular methods for checking the stability of the
superhump period and determining its value is an $O-C$ diagram. Only
data from 2003 were used due to fact that we recorded more subsequent
nights with superhumps in 2003 than in 2005. We analized timings of
primary maxima, because they were almost always high and clearly
detectable in the light curve of the variable star. 
Due to the poor weather conditions on May 10/11, 2003 we obtained short
2.4-hr run with a gap in the midle. Additionaly, the light curve from
that night is characterized by small amplitude variations. Taking into
account these facts we did not estimate the moment of maximum on May
10/11. In total, we were
able to determine 8 moments of maxima, which are listed in Table 2
together with their errors, cycle numbers $E$, and $O-C$ values.\\

\begin{table}[!h]
\caption{Times of maxima in the light curve of V1113 Cyg in May 2003 superoutburst.}
\vspace{0.1cm}
\begin{center}
\begin{tabular}{|r|c|r|r|}
\hline
\hline
$E$ & ${\rm HJD}_{\rm max} - 2452000$ & Error & $O-C$\\
    &                                 &       & [cycles] \\
\hline
0  & 765.440 & 0.001 & $-0.033$\\
1  & 765.521 & 0.001 & $-0.006$\\
13 & 766.473 & 0.001 & $ 0.058$\\
14 & 766.548 & 0.002 & $ 0.085$\\
26 & 767.494 & 0.002 & $-0.003$\\
37 & 768.358 & 0.002 & $-0.054$\\
38 & 768.439 & 0.002 & $-0.027$\\
39 & 768.521 & 0.010 & $ 0.012$\\
\hline
\end{tabular}
\end{center}
\end{table}

The least squares linear fit to the data from Table 2 gives the
following ephemeris for the maxima:

\begin{equation}
{\rm HJD_{\rm max}} = 765.4426(7) + 0.078910(39) \cdot E
\end{equation}

The $O-C$ values corresponding to the ephemeris (1) are shown in Fig. 5
and listed in Table 2.  It seems that the superhump period shows
a decreasing trend. This fact was confirmed by the second-order
polynomial fit to the moment of maxima. The following
ephemeris were estimated: 

\begin{equation}
{\rm HJD_{\rm max}} = 765.4408(7) + 0.07953(13) \cdot E - 1.8(3)\cdot
10^{-5}\cdot E^{2}
\end{equation}

The solid line in the Fig. 5 corresponds to the ephemeris (2).

Finally, we conclude that the period of the superhumps was not stable
during the May 2003 superoutburst of the variable; it can be described
by a decreasing trend with a rate of $\dot P = -4.5(8)\times 10^{-4}$.
Combining both of our superhump period determinations returned its mean
value which is equal to  $P_{\rm sh}=0.07891(3)$ days ($113.63\pm0.04$
min). 

\vspace*{7.2cm}

\includegraphics{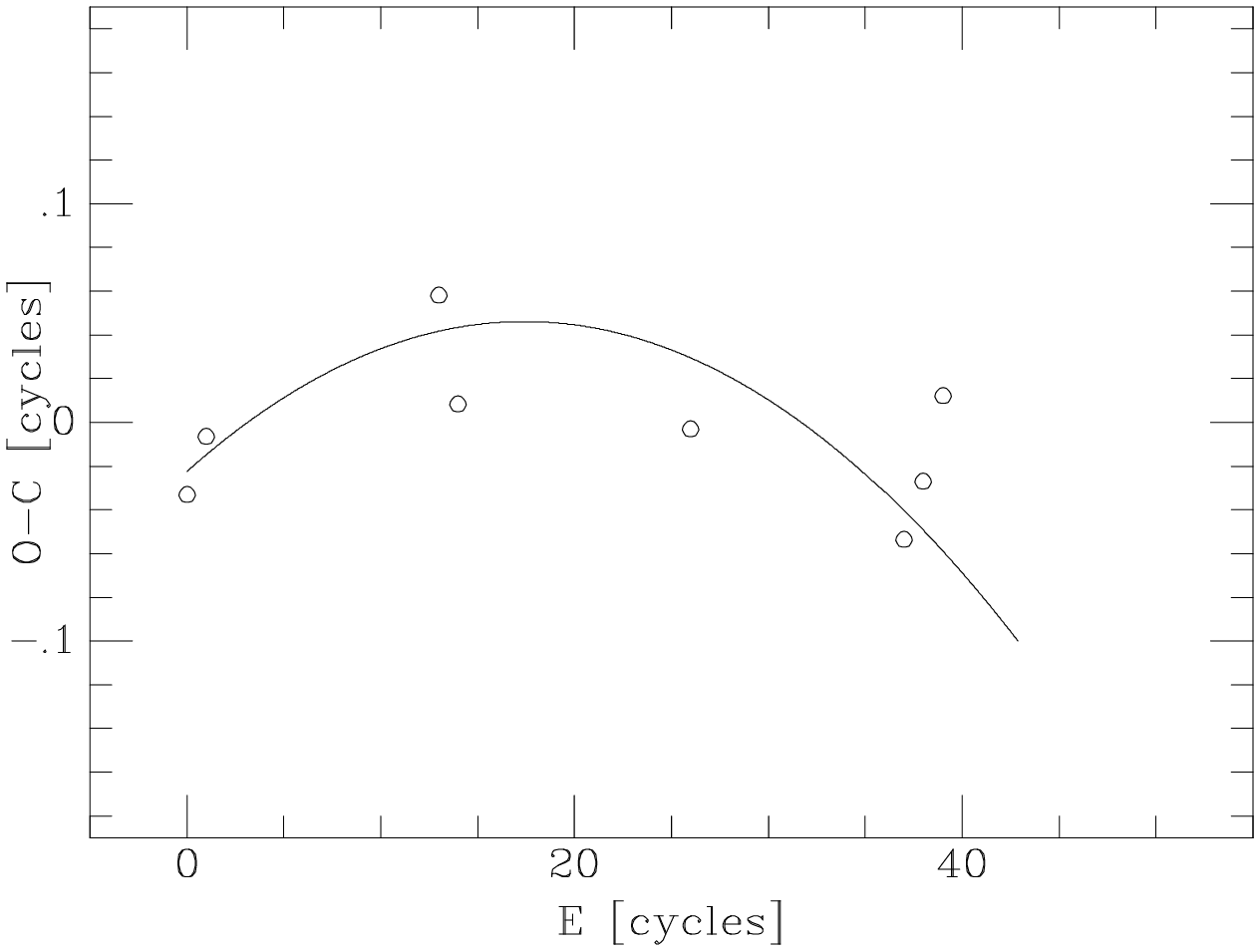}

   \begin{figure}[h]
      \caption {The $O-C$ diagram for superhumps maxima of V1113 Cyg detected in its 
2003 May superoutburst. The solid line corresponds to the fit given by Eq. (2).}
   \end{figure}  

\section{Discussion}

Recently, Kato et al. (2009) published a comprehensive survey of period
variations of superhumps in SU UMa-type dwarf novae. The conlcuded that
in many systems evolution of the superhump period can be divided into
three distinct parts: an early stage with a stable and longer period
(stage A), middle stage for which many systems (especially with $P_{\rm
sh}<0.08$ days) show positive superhump period derivative (stage B) and
final stage with a shorter but again stable superhump period (stage C).
However they reported that V1113 Cyg does not follow this scenario with
decreasing superhump periods with the rates of $-19.2(6.8)\times
10^{-5}$ and $-5.2(4.7)\times 10^{-5}$ observed for
1994 and 2008 superoutbursts, respectively.

V1113 Cyg has its superhump period close to 0.08 day borderline and thus
can show different superhump period evolution. Our high negative $\dot
P$ value may indicate that the behaviour of the star is similar to that
observed in SU UMa or DM Dra (see Figs. 4 and 7 in Kato et al. 2009). 

On the other hand, it is also possible that in 2003 our observations
caught V1113 Cyg during the transition from stage A to B. As can be
estimated from Fig. 4 of Kato et al. (2009), for stars with superhump
period of  around 0.078 days, such a transition occurs around cycle
number $E\approx 20$ assuming $E=0$  for the begining of the
superoutburst (see $O-C$ diagrams of TT Boo, QY Per, RZ Leo or EG Aqr).
The peak of the parabola shown in our Fig. 5 occurs also at $E\approx
20$ but in this case the cycle numeration starts with the first night of
the observations which is not the first night of the superoutburst.
Assuming that the superoutburst started 1-2 days before our first
observation the transition moment shifts to $E\approx 32-33$ or even
$E\approx 45$. It is quite late in comparison to other stars putting
some doubt in A to B transition hypothesis. 

To summarize the results of the observations of the 2003 and 2005
brightenings of V1113 Cyg we can confirm:
\begin{itemize}
\item Detection of the clear superhumps during those
events directly proves that V1113 Cyg belongs to the group of SU UMa
variables. 
\item The amplitude of the superoutburst is at least ($\sim$ 4) mag, 
but we never caught the star at the beginning of the superoutburst. 
According to Kato (2001) this variable has a large ($\sim$ 6) mag outburst 
amplitude and it is a possible estimation.
\item During two years of observation the star went into at least two superoutbursts 
so V1113 Cyg has about one superoutburst a year. Eruptions of the star lasted 
about two weeks and were not shorter than 12 days (July 2005 superoutburst).
\item We did not record any normal outbursts. Lack of  them may be an observational effect. 
But in general we agree with the results of Kato (2001) which suggest low ratio 
of outbursts to superoutbursts.
\item During the first observed superoutburst
the superhump period was decreasing with an enormous rate of $\dot P =
-4.5(8)\times 10^{-4}$ which is one of the highest values ever observed in SU UMa systems.
\end{itemize}

\bigskip \noindent {\bf Acknowledgments.} ~We acknowledge generous
allocation of the Warsaw Observatory 0.6-m telescope time.  Data from
AAVSO observers are also appreciated. We thank to W. Pych for
providing some useful software which was used in the analysis and to
K. Mularczyk and P. K\c{e}dzierski for their assistance in
observations. This work was supported by the Polish MNiSW grant
no. N203~301~335. KZ was supported by Foundation for the Polish
Science through grant MISTRZ.


{\small

\begin{thebibliography}{}
   \bibitem{Hell00} Hellier C., 2001,  {\it Cataclysmic Variable Stars}, Springer.
   \bibitem{Hof01} Hoffmeister C., 1966, Astron. Nachr. 289, 139.
   \bibitem{ka01} Kato T., 2001, Com. 27 and 42 of the IAU, Info. Bull. on Var. Stars, 5110.
   \bibitem{ka02} Kato T. et al., 2003b, MNRAS, 339, 861.
   \bibitem{ka03} Kato T. Imada, A., Uemura, M., et al., 2009, PASJ, 61, S395
   \bibitem{ka04} Kato T., Nogami D., Masuda S., Hirata R., 1996, PASJ, 48, 45.
   \bibitem{Oss02} Osaki Y., 1996, PASP, 108, 39.
   \bibitem{sl07} Schreiber M.R., Lasota J.-P, 2007, A\&A, 473, 897
   \bibitem{sc96} Schwarzenberg-Czerny A., 1996, ApJ Letters,
           460, L107
   \bibitem{sm04} Smak J., 2004, Acta Astron., 54, 221
   \bibitem{sm08} Smak J., 2008, Acta Astron., 58, 55
   \bibitem{sm09} Smak J., 2009, Acta Astron., 59, 121
   \bibitem{st87} Stetson P.B., 1987, PASP, 99, 191.
   \bibitem{up92}  Udalski A., Pych W., 1992, Acta Astron.,
           42, 285.
   \bibitem{War03} Warner B. 1995,  {\it Cataclysmic Variable Stars}, Cambridge University Press.   
   
\end{thebibliography}
}
\end{document}